\documentclass[aps,prd,reprint,preprintnumbers,superscriptaddress,showpacs,twocolumn]{revtex4-1}
\usepackage{latexsym,graphicx,amssymb,amsmath,mathrsfs}
\usepackage{setspace,bm}
\usepackage[breaklinks, colorlinks=true, pdfstartview=FitV, linkcolor=red, citecolor=blue, urlcolor=blue]{hyperref}
\usepackage[usenames]{color}
\usepackage{latexsym}
\usepackage{epstopdf}
\usepackage{mathtools}
\usepackage{url}
\usepackage{comment}
\usepackage{braket}
\usepackage{CJKutf8}

\newcommand{\bequ}{\begin{equation}}
\newcommand{\eequ}{\end{equation}}
\newcommand{\bea}{\begin{eqnarray}}
\newcommand{\eea}{\end{eqnarray}}

\DeclareSymbolFont{boldletters}{OML}{cmm} {b}{it}
\DeclareSymbolFontAlphabet{\mathbit}{boldletters}
\DeclareMathSymbol{\alpha}{\mathalpha}{letters}{"0B}
\DeclareMathSymbol{\beta}{\mathalpha}{letters}{"0C}
\DeclareMathSymbol{\gamma}{\mathalpha}{letters}{"0D}
\DeclareMathSymbol{\delta}{\mathalpha}{letters}{"0E}
\DeclareMathSymbol{\epsilon}{\mathalpha}{letters}{"0F}
\DeclareMathSymbol{\zeta}{\mathalpha}{letters}{"10}
\DeclareMathSymbol{\eta}{\mathalpha}{letters}{"11}
\DeclareMathSymbol{\theta}{\mathalpha}{letters}{"12}
\DeclareMathSymbol{\iota}{\mathalpha}{letters}{"13}
\DeclareMathSymbol{\kappa}{\mathalpha}{letters}{"14}
\DeclareMathSymbol{\lambda}{\mathalpha}{letters}{"15}
\DeclareMathSymbol{\mu}{\mathalpha}{letters}{"16}
\DeclareMathSymbol{\nu}{\mathalpha}{letters}{"17}
\DeclareMathSymbol{\xi}{\mathalpha}{letters}{"18}
\DeclareMathSymbol{\pi}{\mathalpha}{letters}{"19}
\DeclareMathSymbol{\rho}{\mathalpha}{letters}{"1A}
\DeclareMathSymbol{\sigma}{\mathalpha}{letters}{"1B}
\DeclareMathSymbol{\tau}{\mathalpha}{letters}{"1C}
\DeclareMathSymbol{\upsilon}{\mathalpha}{letters}{"1D}
\DeclareMathSymbol{\phi}{\mathalpha}{letters}{"1E}
\DeclareMathSymbol{\chi}{\mathalpha}{letters}{"1F}
\DeclareMathSymbol{\psi}{\mathalpha}{letters}{"20}
\DeclareMathSymbol{\omega}{\mathalpha}{letters}{"21}
\DeclareMathSymbol{\varepsilon}{\mathalpha}{letters}{"22}
\DeclareMathSymbol{\vartheta}{\mathalpha}{letters}{"23}
\DeclareMathSymbol{\varpi}{\mathalpha}{letters}{"24}
\DeclareMathSymbol{\varrho}{\mathalpha}{letters}{"25}
\DeclareMathSymbol{\varsigma}{\mathalpha}{letters}{"26}
\DeclareMathSymbol{\varphi}{\mathalpha}{letters}{"27}
\DeclareMathSymbol{\Gamma}{\mathalpha}{letters}{"00}
\DeclareMathSymbol{\Delta}{\mathalpha}{letters}{"01}
\DeclareMathSymbol{\Theta}{\mathalpha}{letters}{"02}
\DeclareMathSymbol{\Lambda}{\mathalpha}{letters}{"03}
\DeclareMathSymbol{\Xi}{\mathalpha}{letters}{"04}
\DeclareMathSymbol{\Pi}{\mathalpha}{letters}{"05}
\DeclareMathSymbol{\Sigma}{\mathalpha}{letters}{"06}
\DeclareMathSymbol{\Upsilon}{\mathalpha}{letters}{"07}
\DeclareMathSymbol{\Phi}{\mathalpha}{letters}{"08}
\DeclareMathSymbol{\Psi}{\mathalpha}{letters}{"09}
\DeclareMathSymbol{\Omega}{\mathalpha}{letters}{"0A}

\begin{document}
%\preprint{SAGA-HE-???}
\title{Hadron-quark hybrid model, modular transformation\\ and Roberge-Weiss transition}

\author{Hiroaki Kouno}
\email[]{kounoh@cc.saga-u.ac.jp}
\affiliation{Department of Physics, Saga University,
             Saga 840-8502, Japan}

\author{Riki Oshima}
\email[]{24804001@edu.cc.saga-u.ac.jp}
\affiliation{Department of Physics, Saga University,
             Saga 840-8502, Japan}

\author{Motoi Tachibana}
\email[]{motoi@cc.saga-u.ac.jp}
\affiliation{Department of Physics, Saga University,
             Saga 840-8502, Japan}
\affiliation{Center for Theoretical Physics, Khazar University, 41 Mehseti Street, Baku, AZ1096, Azerbaijan}

\author{Kouji Kashiwa}
\email[]{kashiwa@fit.ac.jp}
\affiliation{Fukuoka Institute of Technology, Wajiro, Fukuoka 811-0295, Japan}

%\date{\today}

\begin{abstract}
In the framework of modular transformations, we reformulate the recently proposed hadron-quark hybrid model when the imaginary baryonic chemical potential is introduced. 
In this case,  the number density of the hybrid model is obtained by the modular transformation of the complex number densities of the  baryons (antibaryons) and the quarks (antiquarks). 
We can regard these number densities as the basis in the complex plane. 
As a result, we can consider the torus which is characterized by the basis. 
Since the complex structure of the torus is invariant under the modular transformation, we can extract the topological property of the hadron-quark system using the untransformed baryon (antibaryon) and the quark (antiquark) number densities. 
We apply this model to analyze the Roberge-Weiss transition. 
It is shown that the torus vanishes at the baryonic chemical potential where the Roberge-Weiss transition appears because the number density of baryons (antibaryons) is not linearly independent of the number density of quarks (antiquarks).  
When the temperature $T$ is lower than the Roberge-Weiss transition temperature $T_{\rm RW}$, the torus shrinks smoothly to the one-dimensional object at the Roberge-Weiss transition point, but the discontinuity does not appear. 
On the other hand, the discontinuity of the geometrical object appears when $T>T_{\rm RW}$. 
We also calculate the modulus of the torus and transform it into the fundamental region. 
The transformed moduli are symmetric below $T_{\rm RW}$, but the symmetry is broken above $T_{\rm RW}$. 
\end{abstract}

\maketitle

%%%%%%%%%%%%%%%%%%%%%%%%%%%%%%%%%%%%%%%%%%%%%%%%%%%%%%%%%%%%%%%%%%%%%%%%%%%%%%%%%%%%%%%%%%%%%%%%%%%%%

\section{Introduction}

Determination of the phase diagram of quantum chromodynamics (QCD) is an important subject not only in nuclear and particle physics but also in cosmology and astrophysics; see, e.g., Ref.~\cite{Fukushima:2010bq} and references therein.  
However, when the baryon number chemical potential $\mu$ is finite and real, the fist principle calculation, that is, the lattice QCD (LQCD) simulation, is not feasible due to the infamous sign problem; see Refs.\,\cite{deForcrand:2010ys,Nagata:2021bru,*Nagata:2021ugx} as an example. 
To circumvent the sign problem, several methods are proposed and investigated, although, at present, these methods are not complete and we do not have adequate information on the equation of state (EoS) at finite real baryon density. 
 
One possible way to avoid the sign problem is to use the LQCD results with the imaginary baryon number chemical potential; see Refs.\,\cite{deForcrand:2002ci,D'Elia:2002gd,D'Elia:2004at,Chen:2004tb,D'Elia:2009qz} as an example.
When the baryon number chemical potential $\mu$ is pure imaginary, there is no sign problem. 
One can perform LQCD simulations at finite pure imaginary $\mu$, and then make an analytic continuation from the quantities at imaginary $\mu$ to those at real $\mu$. 
Alternatively, one may determine the unknown parameters of an effective model of QCD using the LQCD results at imaginary $\mu$. 
After determining the parameters, the model calculations can be performed at real $\mu$~\cite{Sakai:2009dv}.  
One can also construct the canonical partition function with fixed baryon number from the grand canonical partition function with the pure imaginary chemical potential ~\cite{Roberge:1986mm}. 
In principle, the grand canonical partition function with the real chemical potential can be constructed from the canonical ones. 
However, at present, these approaches are far from perfection.  
   
\begin{figure}[t]
\centering
\includegraphics[width=0.40\textwidth]{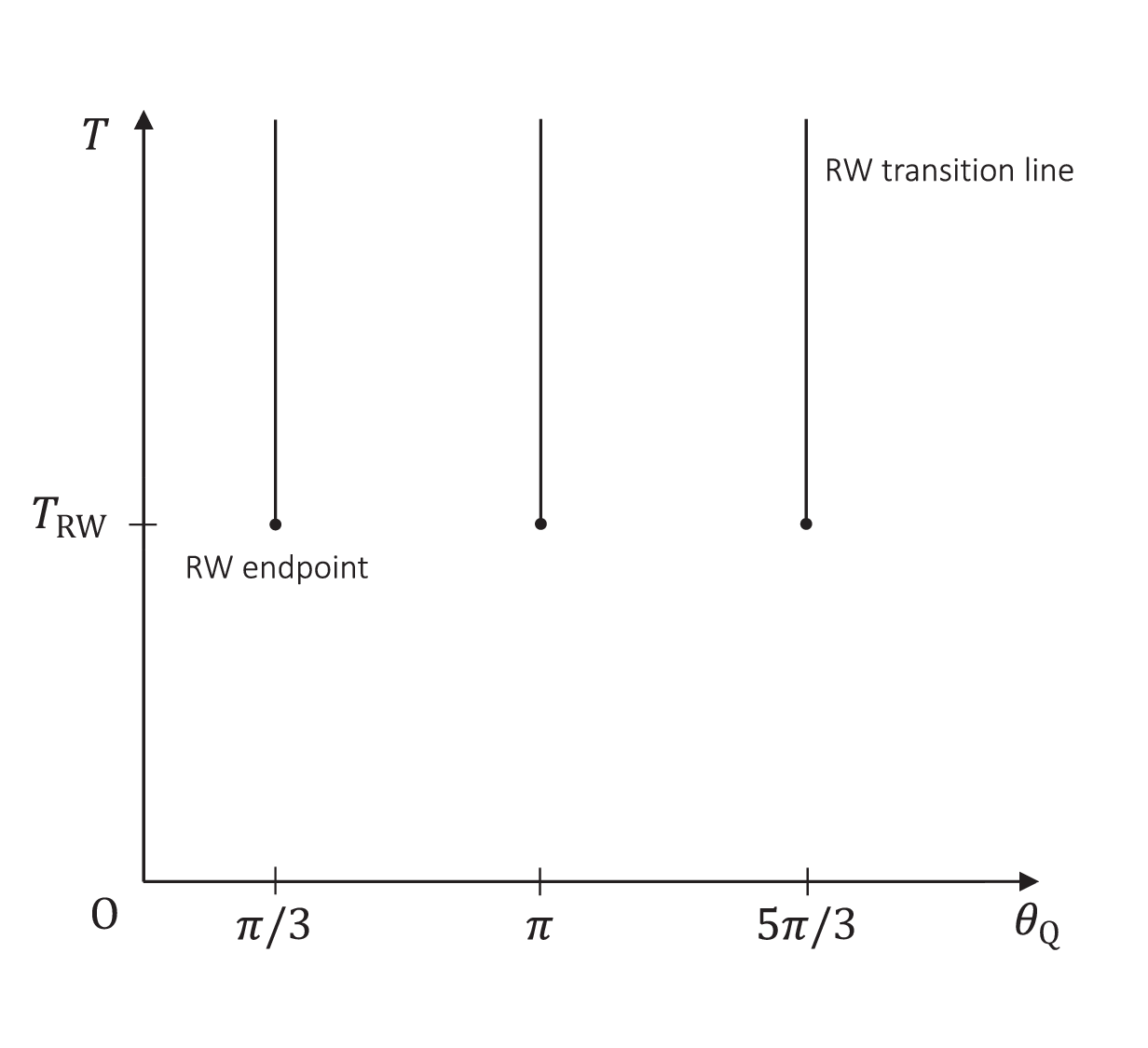}
\caption{The schematic phase diagram of the RW transitions. 
The three solid lines show the RW transition lines.  
The dots at the end of the lines are the RW endpoints.
}
\label{F_phaseD}
\end{figure}

QCD at imaginary $\mu$ itself has very interesting properties and may provide us with an important insight into the physical QCD at real $\mu$.  
The grand canonical partition function $Z(\theta_{\mathrm Q})$ with pure imaginary quark chemical potential ($\mu_{\mathrm Q}={\mu / 3}=i\theta_{\mathrm Q} T$) has the Roberge-Weiss (RW) periodicity~\cite{Roberge:1986mm} as
%%%%%%%%%%%%%%%%%%%%%%
\begin{eqnarray}
Z \Bigl(\theta_{\mathrm Q} + 
       {2\pi\over{3}} \Bigr) 
= Z(\theta_{\mathrm Q}), 
\label{RWP1}
\end{eqnarray}
%%%%%%%%%%%%%%%%%%%%%%
where $T$ is the temperature and $\theta_\mathrm{Q} \in \mathbb{R}$. 
This periodicity is the remnant of the $\mathbb{Z}_3$-symmetry of pure gluon theory. 
At low temperature, $Z(\theta_{\mathrm Q} )$ is a smooth function of $\theta_{\mathrm Q}$. 
However, at high temperature above the RW temperature $T_{\rm RW}$, it has a singularity at $\theta_{\mathrm Q}=(2k+1) \pi / 3$ where $k \in \mathbb{Z}$.   
This singularity is called the RW transition~\cite{Roberge:1986mm}. 
Figure~\ref{F_phaseD} shows a schematic phase diagram of the RW transitions. 
The order of the endpoint of the RW transition line, which is called the RW endpoint, has not yet been determined definitively. 
$T_{\rm RW}$ for 2+1 flavor QCD is estimated as about $200$~MeV by LQCD simulations~\cite{Bonati:2016pwz,Cuteri:2022vwk,Bonati:2018fvg}. 
The RW transition is well described by quark models such as the Polyakov-loop extended Nambu--Jona-Lasinio (PNJL) model~\cite{Sakai:2008py,Kouno:2009bm}. 
However, such a quark model fails to reproduce the EoS of QCD matter at low temperature.   

%%%%%%%%%%%%%%%
%%%%%%%%%%%%%%%

The numerical results obtained by the hadron resonance gas (HRG) model are known to agree well with the LQCD results if $T$ is not so high. 
%It is known that the LQCD results at $\mu = 0$ are in good agreement with those obtained by the hadron resonance gas (HRG) model when the temperature $T$ is not so high.  
Usually, the ideal gas approximation is used for the calculations in the HRG model.  
In the HRG model with pure imaginary baryon number chemical potential $\mu =i\theta T=i3\theta_{\rm Q}T$, the RW periodicity is trivial, since the model has a trivial periodicity
%%%%%%%%%%%%%%%%%%%%%%
\begin{eqnarray}
Z_{\rm HRG}(\theta  +2\pi )=Z_{\rm HRG}(\theta). 
\label{RWP_HRG}
\end{eqnarray}
%%%%%%%%%%%%%%%%%%%%%%
In the case of the free hadron resonance gas model, $Z_{\rm HRG}(\theta )$ is a smooth function of $\theta$ at any temperature. 
Of course, such a simple hadron model cannot reproduce the RW transition.

The RW transition is expected to have a strong relation to the deconfinement transition. 
A construction of the hybrid model which can reproduce the low- and high-temperature properties of LQCD results as well as the RW transition is desirable.  
In this paper, we reformulate the recently proposed hadron-quark hybrid model
~\cite{Kouno:2023ygw} in the framework of the modular transformation when the pure imaginary baryonic chemical potential is introduced. 
The baryonic number denisty $\tilde{n}_\mathrm{b}$ of the hybrid model is given by
%%%%%%%%%%%%%%
\begin{eqnarray}
\tilde{n}_{\rm b} ={n_{\rm b}\over{1+(n_{\rm b}/n_{\rm q})}}. 
\label{n_b_hybrid_intro}
\end{eqnarray}
%%%%%%%%%%%%%%
where $n_\mathrm{b}$ and $n_\mathrm{q}$ are the number density of baryons and a one-third of the number density of quarks, respectively. 
Similarly, the antibaryonic number density of the hybrid model is given using the number densities of antibaryons and antiquarks. 
When $n_\mathrm{q}$ is nonzero, Eq.~(\ref{n_b_hybrid_intro}) can be rewritten as 
%%%%%%%%%%%%%%
\begin{eqnarray}
\tau^\prime \equiv {\tilde{n}_{\rm b}\over{n_\mathrm{q}}}
={\tau\over{\tau+1}},
\label{modular_intro}
\end{eqnarray}
%%%%%%%%%%%%%%
where $\tau \equiv n_\mathrm{b}/n_\mathrm{q}$. 
At the imaginary baryonic chemical potential, $\tau$ is a complex number in general. 
Hence, Eq.~(\ref{modular_intro}) is a kind of modular transformations. 
For detail description of the modular transformation, see Sec.~\ref{Modulartr}.   

We regard $n_\mathrm{b}$ and $n_\mathrm{q}$ as the basis in the complex plane. 
As a result, we can consider the lattice and the torus, which are characterized by the basis, or equivalently, their ratio $\tau$ which is called modulus of the torus.  
For a brief review of the lattice and the torus, see appendix~\ref{Ltorus}.  
Since the complex structure of the torus is invariant under the modular transformation (\ref{modular_intro}), we can extract the topological property of the hadron-quark system using the untransformed baryon (antibaryon) and the quark (antiquark) number densities. 
Using the torus, we qualitatively analyze the RW transition.  
The torus vanish if $n_\mathrm{b}$ and $n_\mathrm{q}$ is not linearly independent.   
In fact, when $T<T_{\rm RW}$, the torus shrinks smoothly to the one-dimensional object at $\theta =(2k+1)\pi~(\theta_\mathrm{Q} =(2k+1)\pi /3)$, but the discontinuity does not appear. 
On the other hand, the discontinuity of the geometrical object appears there when $T>T_{\rm RW}$. 

This paper is organized as follows. 
In Sec.~\ref{RWP}, the RW periodicity and transition are briefly reviewed.  
In Sec.~\ref{Modulartr}, we briefly summarize the modular transformation.   
In Sec. \ref{Hybridmodel}, the hybrid model is reformulated within the framework of modular transformation. 
In Sec. \ref{RWmodular}, the RW transition is qualitatively analyzed using the hybrid model. 
Section \ref{summary} is devoted to the summary and discussions.

%%%%%%%%%%%%%%%%%%%%%%%%%%%%%%%%%%%%%%%%%%%%%
%%%%%%%%%%%%%%%%%%%%%%%%%%%%%%%%%%%%%%%%%%%%%
%%%%%%%%%%%%%%%%%%%%%%%%%%%%%%%%%%%%%%%%%%%%%
\section{Roberge-Weiss periodicity and transition}
\label{RWP}
%%%%%%%%%%%%%%%%%%%%%%%%%%%%%%%%%%%%%%%%%%%%%
%%%%%%%%%%%%%%%%%%%%%%%%%%%%%%%%%%%%%%%%%%%%%
%%%%%%%%%%%%%%%%%%%%%%%%%%%%%%%%%%%%%%%%%%%%%

The grand canonical partition function of QCD with imaginary $\mu_\mathrm{Q} =i\theta_{\rm Q}T$ is given by
%%%%%%%%%%%%%%
\begin{align}
Z(\theta_{\rm Q})
&= \int {\cal D} \psi {\cal D} \bar{\psi} {\cal D} A_\mu e^{-S(\theta_{\rm Q} )}, 
\label{Z-QCD}
\end{align}
%%%%%%%%%%%%%%
where
%%%%%%%%%%%%%%
\begin{align}
S(\theta_{\rm Q})
&=\int_0^\beta dt_\mathrm{E}\int_{-\infty}^\infty d^3x \, {\cal L}(\theta_{\rm Q}),
\label{S-QCD}
\end{align}
%%%%%%%%%%%%%%
with
%%%%%%%%%%%%%%
\begin{align}
{\cal L} (\theta_{\rm Q})
&= \bar{\psi}(\gamma_\mu D_\mu -m_0)\psi 
 -{1\over{4}}F_{\mu\nu}^2-i{\theta_{\rm Q}\over{\beta}}\bar{\psi}\gamma_4\psi ,
\label{L-QCD}
\end{align}
%%%%%%%%%%%%%%
here $\psi$, $A_\mu$, $D_\mu$, $F_{\mu\nu}$ and $m_0$ are the quark field, the gluon field, the covariant derivative, the field strength of gluon field and the current quark mass matrix, respectively, and $\beta = 1/T$. 
The Euclidean notation is used in Eqs. (\ref{Z-QCD}) $\sim$ (\ref{L-QCD}).  

We perform the gauge transformation   
%%%%%%%%%%%%%%%
\begin{eqnarray}
A_\mu&\mapsto& U({\bf x},t_\mathrm{E} )A_\mu U^{-1}({\bf x},t_\mathrm{E} )-{i\over{g}}(\partial_\mu U({\bf x},t_\mathrm{E} ))U^{-1}({\bf x},t_\mathrm{E} ), 
\nonumber\\
\psi &\mapsto & U({\bf x},t_\mathrm{E} )\psi, 
\label{Z3trans}
\end{eqnarray}
%%%%%%%%%%%%%%%
where $g$ is a coupling constant, $U({\bf x},t_\mathrm{E})$ are elements of $SU(3)$ with the temporal boundary condition $U({\bf x},\beta)=z_3U({\bf x},0)$ and 
$z_3=\exp(-i2\pi k/3)$ is a $\mathbb{Z}_3$ element with $k \in \mathbb{Z}$. 
The action $S(\theta_{\rm Q} )$ is invariant under this transformation but the quark boundary condition changes into  
%%%%%%%%%%%%%%%
\begin{eqnarray}
\psi ({\bf x},\beta )=-\exp{\Big(i{2\pi k\over{3}}\Big)}\psi ({\bf x},0). 
\label{quark_boundary}
\end{eqnarray}
%%%%%%%%%%%%%%%
Next we perform the following transformation of quark field; 
%%%%%%%%%%%%%%%
\begin{eqnarray}
\psi \mapsto \exp{\left(-i{2\pi k t_\mathrm{E} \over{3\beta}}\right)}\psi. 
\label{transform}
\end{eqnarray}
%%%%%%%%%%%%%%%
Then, the boundary condition changes into the ordinary anti-periodic one
%%%%%%%%%%%%%%%
\begin{eqnarray}
\psi ({\bf x},\beta )=-\psi ({\bf x},0), 
\label{quark_boundary_anti}
\end{eqnarray}
%%%%%%%%%%%%%%%
but the chemical potential term in the QCD Lagrangian density (\ref{L-QCD}) also changes into 
%%%%%%%%%%%%%%
\begin{align}
-i{(\theta_{\rm Q}+2\pi k/3)\over{\beta}}\bar{\psi}\gamma_4\psi . 
\label{L_chemical}
\end{align}
%%%%%%%%%%%%%%
Hence, we obtain the RW periodicity (\ref{RWP1}). 
Dynamical quarks break the $\mathbb{Z}_3$-symmetry but the RW periodicity appears as a remnant of the $\mathbb{Z}_3$-symmetry~\cite{Roberge:1986mm}. 
In the following, we concentrate our discussions on the region $0\le \theta_{\rm Q} \leq {2\pi\over{3}}~(0\leq \theta \leq 2\pi)$, since the properties of the RW transition in the other regions are similar to those of this region. 

\begin{figure}[t]
\centering
\includegraphics[width=0.40\textwidth]{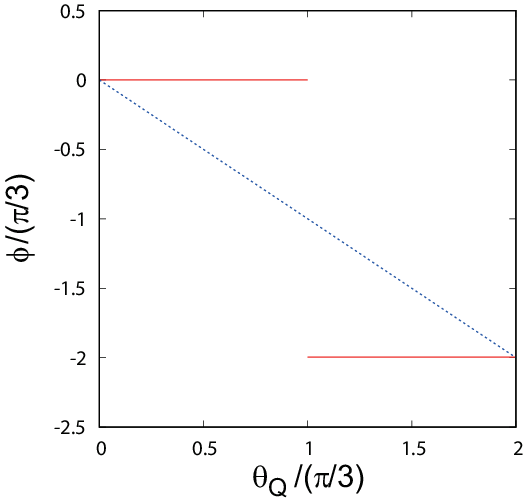}\\
\caption{The approximate $\theta_{\rm Q}$ dependence of the phase $\phi$ of the Polyakov loop $\Phi$. 
The dotted and solid lines show the relations at $T<T_{\rm RW}$ and $T>T_{\rm RW}$, respectively. 
When $T>T_{\rm RW}$, the discontinuity of $\phi$ appears at $\theta_{\rm Q}=\pi/3$.  
}
\label{F_phi_theta}
\end{figure}

At low temperature below the RW transition temperature $T_{\rm RW}$, the phase $\phi$ of the expectation value $\Phi$ of the Polyakov loop is a smooth function of $\theta_{\rm Q}$. 
There is a tendency that $\phi$ cancels the effect of $\theta_{\rm Q}$ and the approximate relation $\phi =-\theta_{\rm Q}$ holds; see Fig.~\ref{F_phi_theta}. 
However, at high temperature above $T_{\rm RW}$, $\phi$ is discontinuous at $\theta_{\rm Q}=\pi/3$ due to the degeneracy of the ground state. 
LQCD simulations indicate the following approximate $\theta_{\rm Q}$-dependence of $\phi$.  
%%%%%%%%%%%%%%%
\begin{eqnarray}
\phi
=
%\left\{
%\begin{array}{cl}
\begin{dcases}
-\theta_{\rm Q} & ~~(T< T_{\rm RW}, ~~0<\theta_{\mathrm Q}<{2\pi\over{3}} ) \\
~~\, 0 & ~~(T>T_{\rm RW},~~~0\leq \theta_{\rm Q}<{\pi\over{3}}) \\
-{2\over{3}}\pi & ~~(T>T_{\rm RW},~~~{\pi\over{3}} < \theta_{\rm Q} \leq {2\pi\over{3}} )
\end{dcases}
%\end{array}
%\right.  . 
.
\label{thetaq_phi}
\end{eqnarray}
%%%%%%%%%%%%%%%
See Fig.~\ref{F_phi_theta} and, also see, e.g., Fig. 7 in Ref.~\cite{Sakai:2009dv} and the references therein. 
%%%%%%%%%%%%%%%
%%%%%%%%%%%%%%%

This relation is also well explained by the effective model of QCD, such as the Polyakov-loop extended Nambu--Jona-Lasinio (PNJL) model
~\cite{Dumitru:2002cf,Fukushima:2003fw,Ratti:2005jh,Ghosh:2006qh,Megias:2004hj,Roessner:2006xn,Kashiwa:2007hw,Sakai:2010rp,Sakai:2009dv,Sakai:2008py,Kouno:2009bm}. 
In the PNJL model, the $\theta_{\rm Q}$-dependence appears only in the form of $\Phi \, e^{i\theta_{\rm Q}}=|\Phi | \, e^{i(\phi +\theta_{\rm Q})}$. 
Hence, at low temperature, $\phi$ tends to cancel the effects of $\theta$ and the approximate relation $\phi = -\theta$ holds. 
At high temperature the $\theta_{\rm Q}$-independent Polyakov-loop potential that originates in the pure gluon sector of QCD has $\mathbb{Z}_3$-symmetric three minima with $\phi =$0, $-{2\pi\over{3}}$ and $-{4\pi\over{3}}$, respectively.   
Due to the effects of the quark part that breaks the $\mathbb{Z}_3$ symmetry explicitly, one of these minima becomes the vacuum of the system.  
When $0\le \theta <{\pi\over{3}}$, $\phi =0$, but $\phi$ jumps from 0 to $-{2\pi\over{3}}$ at $\theta_{\rm Q} ={\pi\over{3}}~(\theta =\pi )$, that is, at the RW transition point.  

It is well known that the PNJL model can reproduce the other several important features of QCD; for example, see Ref.\,\cite{Kashiwa:2019ihm} as a review. 
However, at low temperature, the PNJL model fails to reproduce the EoS of the LQCD simulation.  
Hence, a hybrid model that simultaneously includes the quark and hadron degrees of freedom is needed.

%The RW periodicity and t
%he RW transition are also confirmed by LQCD simulations
%\cite{deForcrand:2002hgr,DElia:2002tig}
%and $T_{\rm RW}$ is estimated as $195\sim 208$~MeV for the 2+1 flavor LQCD simulation~\cite{Bonati:2016pwz,Cuteri:2022vwk,Bonati:2018fvg}. 

%%%%%%%%%%%%%%%%%%%%%%%%%%%%%%%%%%%%%%%%%%%%%%%
%%%%%%%%%%%%%%%%%%%%%%%%%%%%%%%%%%%%%%%%%%%%%%%
%%%%%%%%%%%%%%%%%%%%%%%%%%%%%%%%%%%%%%%%%%%%%%%
\section{Modular transformation}
\label{Modulartr}
%%%%%%%%%%%%%%%%%%%%%%%%%%%%%%%%%%%%%%%%%%%%%%%
%%%%%%%%%%%%%%%%%%%%%%%%%%%%%%%%%%%%%%%%%%%%%%%
%%%%%%%%%%%%%%%%%%%%%%%%%%%%%%%%%%%%%%%%%%%%%%%

In this section, we briefly review the modular transformation.  
The modular transformation is defined by 
%%%%%%%%%%%%%%
\begin{eqnarray}
\tau^\prime =f(\tau )={a\tau +b\over{c\tau +d}}~~~~~(a,b,c,d\in \mathbb{Z}, \Delta =ad-bc=1), 
\nonumber\\
\label{modular}
\end{eqnarray}
%%%%%%%%%%%%%%
where $\tau$ is a complex variable and ${\rm Im} (\tau) \neq 0$.  
If we put $\tau ={z_2\over{z_1}}$ where $z_1$ and $z_2$ are nonzero complex variables,    
Eq. (\ref{modular}) is rewritten as 
%%%%%%%%%%%%%%
\begin{eqnarray}
\tau^\prime
= {z_2^\prime \over{z_1^\prime }}
= {a\tau +b\over{c\tau +d}}
= {a(z_2/z_1)+b\over{c(z_2/z_1)+d}}={az_2+bz_1\over{cz_2+dz_1}}, 
\label{modular_2}
\end{eqnarray}
%%%%%%%%%%%%%%
or 
%%%%%%%%%%%%%%
\begin{eqnarray}
\left( 
\begin{array}{c}
z_2^\prime \\
z_1^\prime
\end{array}
\right)
=
\left( 
\begin{array}{cc}
a &  b   \\
c & d
\end{array}
\right)
\left( 
\begin{array}{c}
z_2 \\
z_1 
\end{array}
\right). 
\label{modular_3}
\end{eqnarray}
%%%%%%%%%%%%%%%
The 2$\times$2 matrix which appears here is an element of $SL(2,\mathbb{Z})$. 
Note that not only $\cal{E}$ but also $-\cal{E}$ is an identity transformation, where $\cal{E}$ is a $2\times 2$ unit matrix.

The modular transformation (\ref{modular}) is generated by the following two transformations and their inverses. 

\bigskip

\noindent
{\bf S transformation};
%%%%%%%%%%%%%%
\begin{align}
&\left( 
\begin{array}{c}
z_2^\prime \\
z_1^\prime
\end{array}
\right)
=
{\cal S}
\left( 
\begin{array}{c}
z_2 \\
z_1 
\end{array}
\right)
\equiv
\left( 
\begin{array}{cc}
0 & -1   \\
1 & 0
\end{array}
\right)
\left( 
\begin{array}{c}
z_2 \\
z_1 
\end{array}
\right)
=
\left( 
\begin{array}{c}
-z_1 \\
z_2 
\end{array}
\right),
\label{modular_S} 
\end{align}
%%%%%%%%%%%%%%%
with $\tau^\prime =-{1\over{\tau}}$. \\

\noindent
{\bf T transformation};
%%%%%%%%%%%%%%
\begin{eqnarray}
\left( 
\begin{array}{c}
z_2^\prime \\
z_1^\prime
\end{array}
\right)
=
\cal{T}
\left( 
\begin{array}{c}
z_2 \\
z_1 
\end{array}
\right)
\equiv
\left( 
\begin{array}{cc}
1 & 1   \\
0 & 1
\end{array}
\right)
\left( 
\begin{array}{c}
z_2 \\
z_1 
\end{array}
\right)
=
\left( 
\begin{array}{c}
z_2 +z_1 \\
z_1 
\end{array}
\right), 
\nonumber\\
\label{modular_T}
\end{eqnarray}
with $\tau^\prime =\tau+1$.
%%%%%%%%%%%%%%%
\bigskip

Note that the S transformation changes the sign of ${\rm Re} (\tau)$ but does not change the sign of ${\rm Im}(\tau )$. 
The T transformation changes ${\rm Re}(\tau )$ but does not ${\rm Im}(\tau )$.  
Consequently, ${\rm Im}(\tau^\prime )>0~(<0)$ when ${\rm Im}(\tau )>0~(<0)$.
The modular transformation does not change the sign of ${\rm Im}(\tau )$. 

If $z_1$ and $z_2$ are linearly independent in the complex plane, we can consider a two dimensional lattice $\Lambda$
%%%%%%%%%%%%%
\begin{eqnarray}
\Lambda =  \{ n_1z_1+n_2z_2 \in \mathbb{C} | n_1,n_2 \in \mathbb{Z} \}, 
\label{lattice}
\end{eqnarray}
%%%%%%%%%%%%%
in the complex plane.  
That is, $z_1$ and $z_2$ are the basis of this lattice. 
We can consider the torus defined by the quotient space $\mathbb{C}/\Lambda$. 
$\tau$ is called a "modulus" of the torus. 
See Appendix~\ref{Ltorus}. 
The modular transformation (\ref{modular_3}) is a base conversion, but this conversion does not change the similarity of lattice and, hence, the complex structures of the torus. 

The modulus of the torus has ambiguity, since the modular transformation does not change the complex structure of the torus. 
However, it is known that $\tau$ can be transformed into the following fundamental region $D_{\rm up}$ by the series of S and T transformations (hence, by the modular transformation), when ${\rm Im}(\tau )>0$;
%%%%%%%%%%%%%%%%%%%%
\begin{eqnarray}
D_{\rm up} 
=\left\{\tau  \Big|  -{1\over{2}}\leq {\rm Re}(\tau) \leq {1\over{2}}, |\tau |\geq 1, {\rm Im}(\tau )>0 \right\}. 
\nonumber
\label{fundamental}
\end{eqnarray}
%%%%%%%%%%%%%%%%%%%%
Note that the left boundary of $D_{\rm up}$ can be changed into the right boundary by the T transformation. 
For a given $\tau$, the transformed modulus is uniquely determined when $|\tau | \neq 1$. 
As is shown in Sec. \ref{RWmodular}, $\tau$ is transformed into the boundary of $D_{\rm up}$ when $|\tau |=1$. 
Similarly,  when ${\rm Im}(\tau )<0$, $\tau$ can be transformed into the fundamental region
%%%%%%%%%%%%%%%%%%%%
\begin{eqnarray}
D_{\rm low}=\left\{\tau  \Big|  -{1\over{2}}\leq {\rm Re}(\tau) \leq {1\over{2}}, |\tau |\geq 1, {\rm Im}(\tau )<0 \right\}, 
\nonumber
\label{fundamental_lower}
\end{eqnarray}
%%%%%%%%%%%%%%%%%%%
by the modular transformation. 
Note that the modular transformation cannot transform $\tau$ in the lower half plane into the one in the upper half plane. 

%%%%%%%%%%%%%%%%%%%%%%%%%%%%%%%%%%%
%%%%%%%%%%%%%%%%%%%%%%%%%%%%%%%%%%%
%%%%%%%%%%%%%%%%%%%%%%%%%%%%%%%%%%%
\section{Hadron-quark hybrid model}
\label{Hybridmodel}
%%%%%%%%%%%%%%%%%%%%%%%%%%%%%%%%%%%
%%%%%%%%%%%%%%%%%%%%%%%%%%%%%%%%%%%
%%%%%%%%%%%%%%%%%%%%%%%%%%%%%%%%%%%

First, we briefly review the hybrid model recently proposed in Ref~\cite{Kouno:2023ygw}.  
Suppose $n_{\rm b}$ and $n_{\rm a}$ are the number densities of point-like baryons and antibaryons, respectively. We can effectively  include the repulsive interaction among baryons (antibaryons) in the model by considering the excluded volume effects (EVE).  
For the EVE, see, e.g., Ref.~\cite{Kouno:2023ygw} and the references therein. 
The modified number density  $\tilde{n}_{\rm b}$ ($\tilde{n}_{\rm a}$) of baryons (antibaryons) is given by 
%%%%%%%%%%%%%%
\begin{eqnarray}
\tilde{n}_{\rm b} ={n_{\rm b}\over{1+vn_{\rm b}}},~~~~~
\tilde{n}_{\rm a} ={n_{\rm a}\over{1+vn_{\rm a}}},  
\label{n_b}
\end{eqnarray}
%%%%%%%%%%%%%%
where $v$ is the volume of a baryon (an antibaryon). 
The net baryon number density is given by $n_{\rm b}-n_{\rm a}$. 
Other thermodynamic quantities such as the pressure and the energy density can be obtained by using thermodynamic relations.  
However, when $n_{\rm b}~(n_{\rm a})\to \infty$, $\tilde{n}_{\rm b}$ ($\tilde{n}_{\rm a}$) approaches the constant value $1/v$, the EoS of baryon matter becomes very hard and the speed of sound exceeds 1 ~\cite{Kouno:2023ygw}.   
The causality is easily violated.  
Hence, the $T$ and/or $\mu$ dependence of $v$ is very important.

Hereafter, we concentrate our discussion on baryons, since the discussion on antibaryons is similar to that on baryons. 
Since the baryon consists of three quarks, we put $v=1/n_{\rm q}$ where $n_{\rm q}$ is a one-third of the number density of the quark in a pure quark model such as the PNJL model.  
The number density of baryons is  given by 
%%%%%%%%%%%%%%
\begin{eqnarray}
\tilde{n}_{\rm b} ={n_{\rm b}\over{1+(n_{\rm b}/n_{\rm q})}}. 
\label{n_b_hybrid}
\end{eqnarray}
%%%%%%%%%%%%%%
If we use the HRG model for the calculation of $n_{\rm b}$, $n_{\rm b}/n_{\rm q}$ becomes much larger than 1 when $T$ or $\mu$ is large. 
Hence, $\tilde{n}_{\rm b}$ approaches $n_{\rm q}$.  
Then, the system can be regarded as quark matter.    
In this way, the natural hadron-quark hybrid model is obtained. 
When $n_{\rm b}$ is large, finite size baryons merge each other and form quark matter. 

It should be remarked that the relation $v=1/n_{\rm q}$ is not valid in the limit $T, \mu \to 0$ where $n_{\rm q}$ also approaches $0$. 
It is known that nucleon has a finite volume $v_0={4\pi\over{3}}r_0^3$ with $r_0\sim 0.8$~fm in vacuum.   
In Ref.~\cite{Kouno:2023ygw} $v$ which interpolates $1/n_{\rm q}$ and $v_0$ is used. 
However, in this paper, we restrict our discussion to the region where $T$ is above or just below $T_{\rm RW}$.  
Hence, we use the simple relation $v=1/n_{\rm q}$ in this paper.

Putting $\tau ={n}_{\rm b}/n_{\rm q}$, we define 
%%%%%%%%%%%%%%
\begin{align}
\tau_{\mathrm h}
 \equiv {\tilde{n}_{\rm b}\over{n_{\rm q}}}
&={n_{\rm b}/n_{\rm q}\over{(n_{\rm b}/n_{\rm q})+1}}
={\tau\over{\tau +1}}
\nonumber\\
&={n_{\rm b}\over{n_{\rm b}+n_{\rm q}}}
 = {n_{\rm b}^\prime\over{n_{\rm q}^\prime}}, 
\label{n_b_hybrid_2} 
\end{align}
%%%%%%%%%%%%%%
where $n_{\rm q}^\prime = n_{\rm q}+n_{\rm b}$ and $n_{\rm b}^\prime =n_{\rm b}$. When $\mu$ is imaginary, $n_{\rm b}$ and $n_{\rm q}$ are complex in general. 
Hence, Eq.~(\ref{n_b_hybrid_2}) is nothing but a modular transformation (\ref{modular}) with $a=c=d=1$ and $b=0$. 
Note that this transformation is induced by the matrix product $\cal{TST}$. 
In fact, we can rewrite (\ref{n_b_hybrid_2}) as 
%%%%%%%%%%%%%%
\begin{eqnarray}
\tau_{\mathrm h} ={-1\over{\tau +1}}+1. 
\label{n_b_hybrid_3}
\end{eqnarray}
%%%%%%%%%%%%%%%%
Hence, we obtain the following conclusions.

\begin{figure}[t]
\centering
\includegraphics[width=0.40\textwidth]{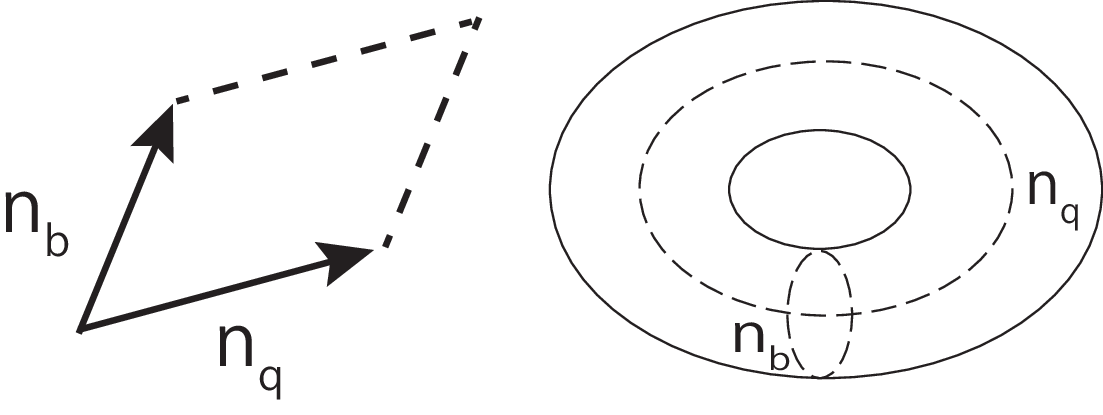}\\

\caption{The unit lattice and the torus. 
}
\label{F_torus}
\end{figure}

\noindent
(1) When $n_{\rm b}$ and $n_{\rm q}$ are linearly independent, we can consider the two dimensional lattice and the corresponding torus which is characterized by $n_{\rm q}$ and $n_{\rm b}$. 
See Fig.~\ref{F_torus}. 

\noindent
(2) Equation (\ref{n_b_hybrid_2}) does not change the complex structure of the torus. 
Hence, only the information of the untransformed quantities $n_{\rm q}$ and $n_{\rm b}$ is needed to analyze the complex structure of the torus.  

It should be remarked that we cannot consider the lattice and torus if $n_{\rm b}$ and $n_{\rm q}$ are not linearly independent. 
In the next section, we show that this case really occurs at the RW transition point.  

%%%%%%%%%%%%%%%
%%%%%%%%%%%%%%%

%%%%%%%%%%%%%%%%%%%%%%%%%%%%%%%%%%%%%%%%%%%%%%%%%%%%%%%%%%%%%%%%%%%%%%%%%%
%%%%%%%%%%%%%%%%%%%%%%%%%%%%%%%%%%%%%%%%%%%%%%%%%%%%%%%%%%%%%%%%%%%%%%%%%%
%%%%%%%%%%%%%%%%%%%%%%%%%%%%%%%%%%%%%%%%%%%%%%%%%%%%%%%%%%%%%%%%%%%%%%%%%%
\section{Qualitative analyses of the Roberge-Weiss transition based on modular transformation}
\label{RWmodular}
%%%%%%%%%%%%%%%%%%%%%%%%%%%%%%%%%%%%%%%%%%%%%%%%%%%%%%%%%%%%%%%%%%%%%%%%%%
%%%%%%%%%%%%%%%%%%%%%%%%%%%%%%%%%%%%%%%%%%%%%%%%%%%%%%%%%%%%%%%%%%%%%%%%%%
%%%%%%%%%%%%%%%%%%%%%%%%%%%%%%%%%%%%%%%%%%%%%%%%%%%%%%%%%%%%%%%%%%%%%%%%%%

In this section, we analyze the Roberge-Weiss transition qualitatively by using the hybrid model based on the modular transformation. 
We use the HRG model for the hadron model and the PNJL model for the quark model with the Boltzmann approximation, which is a good approximation when $|{\rm Re} (\mu ) |$ is not so large. 
We concentrate our discussion on baryons and quarks, since the discussions on antibaryons and antiquarks are similar to those on baryons and quarks. 
For simplicity of the notation in this section, we use $\mu$ and $\theta$ instead of $\mu_{\rm q}$ and $\theta_{\rm Q}$.  

\subsection{Vanishing of the torus}

When the Boltzmann approximation is used, the number density $n_{\rm b}$ of baryons in HRG model is given by 
%%%%%%%%%%%%%%
\begin{eqnarray}
n_{\rm b} &=& B(T) \, e^{\mu /T} \equiv \left(\sum_iB_i(T)\right) e^{\mu /T},
\label{n_HRG}
\end{eqnarray}
%%%%%%%%%%%%%%%
where
%%%%%%%%%%%%%%
\begin{eqnarray}
B_i(T)&=&{g_{s,i}\over{2\pi^2}}\int_0^\infty dp \, p^2e^{-\sqrt{p^2+M_i^2}/T}, 
\label{BT_i}
\end{eqnarray}
%%%%%%%%%%%%%%%
here $g_{s,i}$ and $M_i$ are the spin degree of freedom and mass of $i$-th baryon, respectively. 
Under the same approximation, the one-third of number density of quarks in PNJL model is given by
%%%%%%%%%%%%%%
\begin{align}
n_{\rm q}
&= Q(T) \, \Phi \, e^{\mu /(3T)}
\nonumber\\
&\equiv \left(\sum_f Q_f(T)\right)\Phi e^{\mu /(3T)},
\label{n_PNJL}
\end{align}
%%%%%%%%%%%%%%%
where
%%%%%%%%%%%%%%
\begin{align}
Q_f(T)
&= {g_{s,f}\over{2\pi^2}}\int_0^\infty dp \, p^2e^{-\sqrt{p^2+M_f^2}/T}, 
\label{QT_f}
\end{align}
%%%%%%%%%%%%%%%
here $g_{s,f}$ and $M_f$ are the spin degree of freedom and mass of the flavor $f$ quarks, respectively, and $\Phi$ is the averaged value of the (traced) Polyakov loop.

The (untransformed) modulus of the torus constructed from $n_{\rm q}$ and $n_{\rm b}$ is given by 
%%%%%%%%%%%%%%%
\begin{eqnarray}
\tau =|\tau | \, e^{i\Delta \phi_{\mathrm b-q}}={n_{\rm b}\over{n_{\rm q}}}={B(T)\over{Q(T)|\Phi |}}e^{i(\phi_{\rm b}-\phi_{\rm q})}, 
\label{modular_QCD}
\end{eqnarray}
%%%%%%%%%%%%%%%
where $\phi_{\rm b}$ and $\phi_{\rm q}$ are the phases of $n_{\rm b}$ and $n_{\rm q}$, respectively. 

When $\mu =i\theta T~(0\leq \theta <2\pi)$, 
the phase $\phi_{\rm b}$ of $n_{\rm b}$ is given by
%%%%%%%%%%%%%%%
\begin{eqnarray}
\phi_{\rm b}=\theta, 
\label{phase_b}
\end{eqnarray}
%%%%%%%%%%%%%%%%
where we have omitted the trivial ambiguity $2 n\pi$ with an arbitrary integer $n$.

As was mentioned in Sec.~\ref{RWP},  
at low temperature below $T_{\rm RW}$, there is a tendency that the phase $\phi$ of $\Phi$ cancels the effect of $\theta$.  
Hence, the phase $\phi_{\rm q}$ of $n_{\rm q}$ is approximately given by  
%%%%%%%%%%%%%%%
\begin{eqnarray}
\phi_{\rm q}= {\theta\over{3}}+{\phi} =0. 
\label{phase_q_low}
\end{eqnarray}
%%%%%%%%%%%%%%%
In particular, it seems that Eq. (\ref{phase_q_low}) holds exactly when $\theta =0,\pi$ in LQCD simulation. 
See, e.g., Fig. 7 in Ref~\cite{Sakai:2009dv}. and references therein. 
The phase of the modulus $\tau$ is given by
%%%%%%%%%%%%%%%
\begin{eqnarray}
\Delta \phi_{\mathrm b-q} =\phi_{\rm b}-\phi_{\rm q}=\theta .
\label{D_phase_low}
\end{eqnarray}
%%%%%%%%%%%%%%%
At $\theta =\pi$, $n_{\rm b}$ is antiparallel to $n_{\rm q}$, and, hence, it is not linearly independent with $n_{\rm q}$.   
There, the torus shrinks smoothly to a one-dimensional object.  
Although there is no discontinuity of $\Delta \phi_{\mathrm b-q}$ below $T_{\rm RW}$, in this sense, the point $\theta =\pi$ is special. 
Of course, $n_{\rm b}$ is parallel to $n_{\rm q}$ also at $\theta =0$, but this fact is trivial since $\theta$ itself vanishes.

As was also mentioned in Sec.~\ref{RWP}, at high temperature above $T_{\rm RW}$, $\phi$ is given by 
%%%%%%%%%%%%%%%
\begin{eqnarray}
\phi
=
%\left\{
%\begin{array}{cl}
\begin{dcases}
~~\, 0 & ~~(0 \leq \theta < \pi) \\
-{2\over{3}}\pi & ~~(\pi < \theta \leq 2\pi )
\end{dcases}
%\end{array}
%\right. 
.
\label{phi_high}
\end{eqnarray}
%%%%%%%%%%%%%%%
Note that two vacua with $\phi =0$ and $-2\pi/3$ are degenerated at $\theta =\pi$. 
The phase $\phi_{\rm q}$ is given by
%%%%%%%%%%%%%%%
\begin{eqnarray}
\phi_{\rm q} 
=
%\left\{
%\begin{array}{cl}
\begin{dcases}
~~~~\, {1\over{3}}\theta & ~~ (0\leq \theta <\pi ) \\
{1\over{3}}\theta-{2\over{3}}\pi & ~~(\pi<\theta \leq 2\pi )
\end{dcases}
.
%\end{array}
%\right. .  
\label{phase_q_high}
\end{eqnarray}
%%%%%%%%%%%%%%%
Hence, the phase of $\tau$ is given by
%%%%%%%%%%%%%%%%
\begin{eqnarray}
\Delta \phi_{\mathrm b-q}  
=\phi_{\rm b}-\phi_{\rm q}
=
%\left\{
%\begin{array}{cl}
\begin{dcases}
~~~\, {2\over{3}}\theta & ~~(0\leq \theta <\pi ) \\
{2\over{3}}\theta + {2\over{3}}\pi & ~~(\pi <\theta \leq 2\pi )
\end{dcases}
.
%\end{array}
%\right. .
\label{D_phase_high}
\end{eqnarray}
%%%%%%%%%%%%%%%
We see that $\Delta \phi_{\mathrm b-q}$ is not continuous at $\theta =\pi$, and, hence the correspondent torus does not. 

%%%%%%%%%%%%%%%%%%%%%%%%%%%%%%%%%%%%%%%%%%
\subsection{Calculation of moduli in the fundamental region} 
%%%%%%%%%%%%%%%%%%%%%%%%%%%%%%%%%%%%%%%%%%

To proceed the analyses further, we use the following approximation and assumption. 

~

\noindent
(1) In PNJL model, $Q(T)$ depends on $\theta$, since the chiral condensates depend on $\theta$. 
$|\Phi |$ also depends on $\theta$. 
We neglect these $\theta$-dependences. 
In this approximation, $|\tau |$ does not depend on $\theta$ either. 
This approximation is valid since our main interests are the behavior of $\tau$ near the RW transition point $\theta =\pi$. 

~

\noindent 
(2) As is seen in (\ref{n_b_hybrid}), $\tilde{n}_{\rm b}\sim n_{\rm b}$ when $|n_{\rm b}/n_{\rm q}| \ll 1$, while $\tilde{n}_{\rm b}\sim n_{\rm q}$ when $|n_{\rm b}/n_{\rm q}| \gg 1$.  
Hence, we assume $|\tau | <1$ when $T<T_{\rm RW}$, and $|\tau |>1$ when $T>T_{\rm RW}$. 

Under these approximations and assumptions, for the fixed value of $|\tau |$, we calculate $\tau$ as a function of $\theta$, then transform it into the fundamental region $D_{\rm up}$ or $D_{\rm low}$. 
We denote the transformed $\tau$ by $\tau_{\rm f}$. 
Note that ${\rm Im}(\tau ) >0$ when $0<\theta <\pi$, while ${\rm Im}(\tau ) <0$ when $\pi <\theta <2\pi$. 
$\tau$ is transformed into $D_{\rm up}$ ($D_{\rm low}$) when $0<\theta <\pi$ ($\pi <\theta <2\pi$). 
Since the transformation into the fundamental region is not continuous transformation, we calculate about 7200$\sim$72000 points for each figure and denote them by dots. Note that the dots often seem to form the solid line. 

\begin{figure}[t]
\centering
\includegraphics[width=0.4\textwidth]{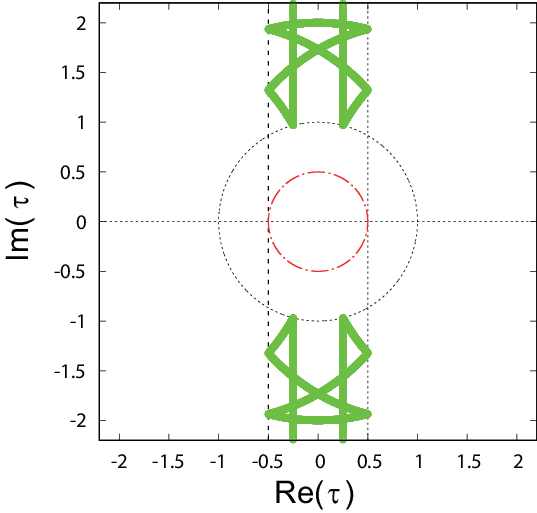}\\
\caption{$\tau$ and $\tau_{\rm f}$ when $T<T_{\rm RW}$ and $|\tau |=0.5$. 
The dot-dashed line and the dots are $\tau$ and $\tau_{\rm f}$, respectively. 
$\tau (\theta \to 0)=(0.5, 0)$, $\tau (\theta\to \pi \pm 0)=(-0.5, 0)$ and $\tau (\theta\to 2\pi)=(0.5, 0)$, while the absolute values of  
$\tau_{\rm f} (\theta \to 0)$, $\tau_{\rm f} (\theta\to \pi \pm 0)$ and $\tau_{\rm f} (\theta\to 2\pi)$ are large. 
}
\label{F_r05_l}
\end{figure}

Figure~\ref{F_r05_l} shows the moduli $\tau$ and $\tau_{\rm f}$ when $T<T_{\rm RW}$ and $|\tau |=0.5$. 
$\tau$ and $\tau_{\rm f}$ are symmetric with respect to the line ${\rm Re}(\tau )=0$. 
There is a left-right symmetry. 
Although the torus vanishes at $\theta=\pi$, $\tau$ is continuous there. 
%%%%%%%%%%%%%
%%%%%%%%%%%%%%%

%%%%%%%%%%%%%
\begin{figure}[t]
\centering
\includegraphics[width=0.4\textwidth]{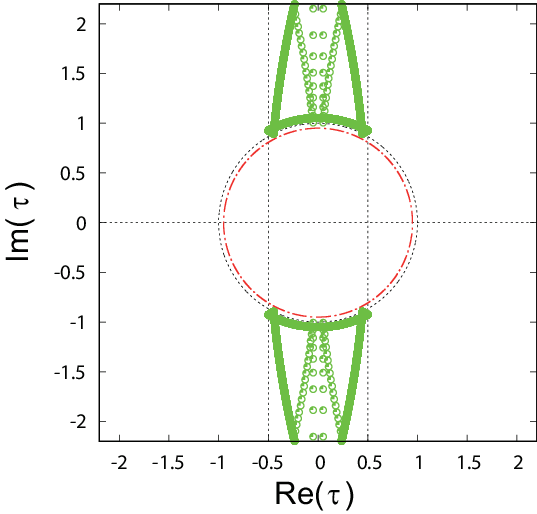}\\
\caption{$\tau$ and $\tau_{\rm f}$ when $T<T_{\rm RW}$ and $|\tau |=0.95$. 
The dot-dashed line and the dots are $\tau$ and $\tau_{\rm f}$, respectively. 
$\tau (\theta \to 0)=(0.95, 0)$, $\tau (\theta\to \pi \pm 0)=(-0.95,0)$ and $\tau (\theta \to 2\pi )=(0.95,0)$ while the absolute values of 
$\tau_{\rm f} (\theta \to 0)$, $\tau_{\rm f} (\theta\to \pi \pm 0)$ and $\tau_{\rm f} (\theta\to 2\pi )$ are large. 
}
\label{F_r095_l}
\end{figure}
%%%%%%%%%%%%%%%

%%%%%%%%%%%%%
\begin{figure}[t]
\centering
\includegraphics[width=0.4\textwidth]{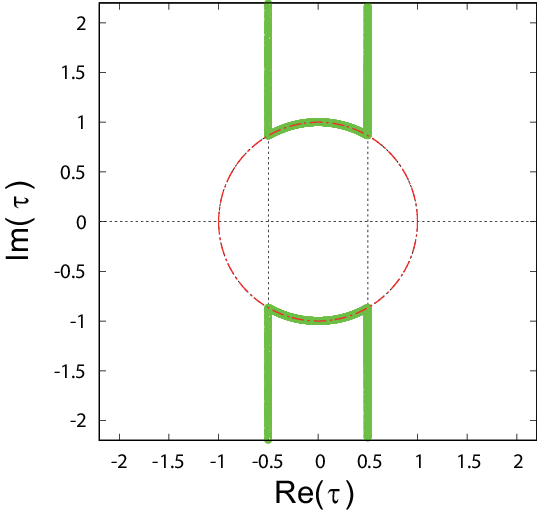}\\
\caption{$\tau$ and $\tau_{\rm f}$ when $T\to T_{\rm RW}-0$ and $|\tau |=1$. 
The dot-dashed line and the dots are $\tau$ and $\tau_{\rm f}$, respectively.  
$\tau (\theta \to 0)=(1.0, 0)$, $\tau (\theta\to \pi \pm 0)=(-1.0,0)$ and $\tau (\theta \to 2\pi )=(1.0,0)$ while   
$\tau_{\rm f} (\theta \to 0)$, $\tau_{\rm f} (\theta\to \pi \pm 0)$ and $\tau_{\rm f} (\theta\to 2\pi )$ show divergent behaviors.  
}
\label{F_r10_l}
\end{figure}
%%%%%%%%%%%%%%%

%%%%%%%%%%%%%
\begin{figure}[t]
\centering
\includegraphics[width=0.4\textwidth]{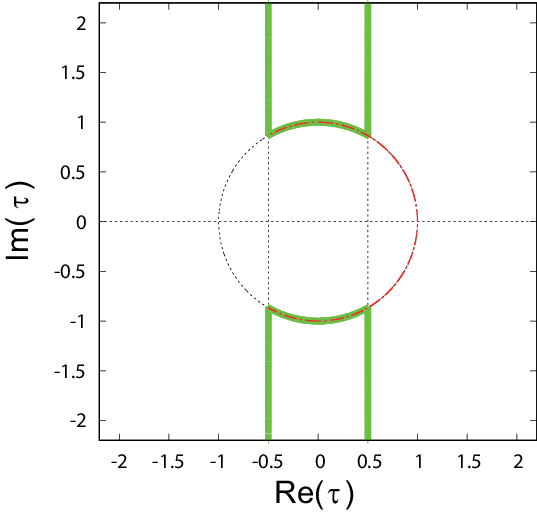}\\
\caption{$\tau$ and $\tau_{\rm f}$ when $T\to T_{\rm RW}+0$ and $|\tau |=1$. 
The dot-dashed line and the dots are $\tau$ and $\tau_{\rm f}$, respectively. 
$\tau (\theta \to 0)=(1.0, 0)$, $\tau (\theta\to \pi - 0)=(-0.5, \sqrt{3}/2)$, $\tau (\theta\to \pi + 0)=(-0.5, -\sqrt{3}/2)$, $\tau (\theta \to 2\pi )=(1.0, 0)$, $\tau_{\rm f} (\theta\to \pi - 0)=(\pm 0.5, \sqrt{3}/2)$ and $\tau_{\rm f} (\theta\to \pi + 0)=(\pm0.5, -\sqrt{3}/2)$ while $\tau_{\rm f} (\theta \to 0)$ and $\tau_{\rm f} (\theta\to 2\pi )$ show divergent behaviors.  
}
\label{F_r10_h}
\end{figure}
%%%%%%%%%%%%%%%

%%%%%%%%%%%%%
\begin{figure}[t]
\centering
\includegraphics[width=0.4\textwidth]{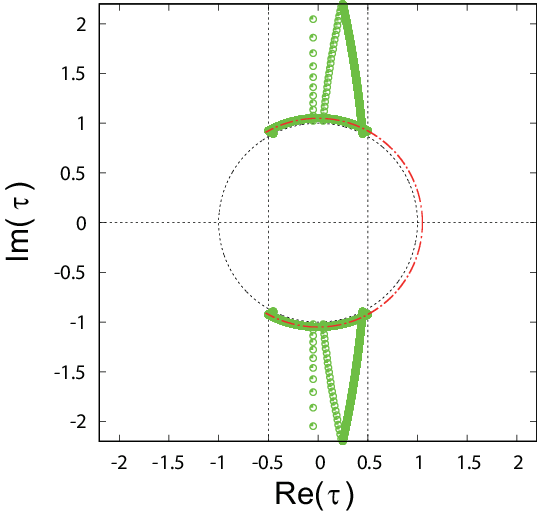}\\
\caption{$\tau$ and $\tau_{\rm f}$ when $T>T_{\rm RW}$ and $|\tau |=1.05$. 
The dot-dashed line and the dots are $\tau$ and $\tau_{\rm f}$, respectively. 
$\tau (\theta \to 0)=(1.05,0)$, $\tau (\theta \to \pi -0)=(-1.05/2,1.05\sqrt{3}/2)$, 
$\tau (\theta \to \pi +0)=(-1.05/2,-1.05\sqrt{3}/2)$, 
$\tau (\theta \to 2\pi )=(1.05,0)$, $\tau_{\rm f} (\theta \to \pi -0)=(0.475,0.909)$ and  $\tau_{\rm f} (\theta \to \pi + 0)=(0.475,-0.909)$ while the absolute values of  
$\tau_{\rm f} (\theta \to 0)$ and $\tau_{\rm f} (\theta \to 2\pi )$ are large.   
}
\label{F_r105_h}
\end{figure}
%%%%%%%%%%%%%%%

%%%%%%%%%%%%%
\begin{figure}[t]
\centering
\includegraphics[width=0.4\textwidth]{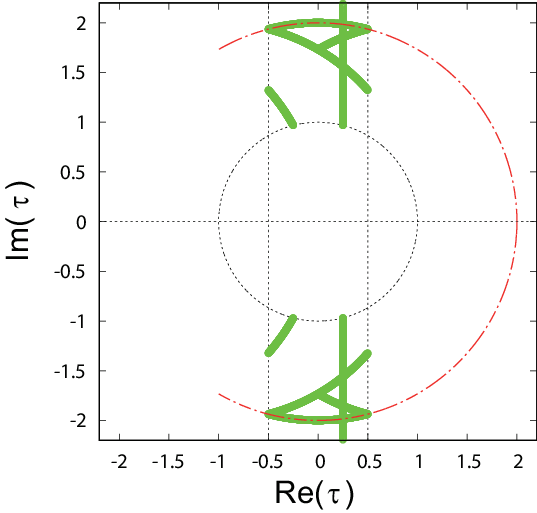}\\
\caption{$\tau$ and $\tau_{\rm f}$ when  $T>T_{\rm RW}$ and $|\tau |=2$. 
The dot-dashed line and the dots are $\tau$ and $\tau_{\rm f}$, respectively. 
$\tau (\theta \to 0)=(2.0,0)$, $\tau (\theta \to \pi -0)=(-1.0,\sqrt{3})$, $\tau (\theta \to \pi +0)=(-1.0,-\sqrt{3})$, 
$\tau_{\rm f} (\theta \to \pi -0)=(0,\sqrt{3})$ and  $\tau_{\rm f} (\theta \to \pi + 0)=(0,-\sqrt{3})$ while the absolute values of  
$\tau_{\rm f} (\theta \to 0)$ and $\tau_{\rm f} (\theta \to 2\pi )$ are large. 
}
\label{F_r20_h}
\end{figure}
%%%%%%%%%%%%%%%

%%%%%%%%%%%%%%%
\begin{figure}[t]
\centering
\bigskip
\bigskip
\includegraphics[width=0.27\textwidth]{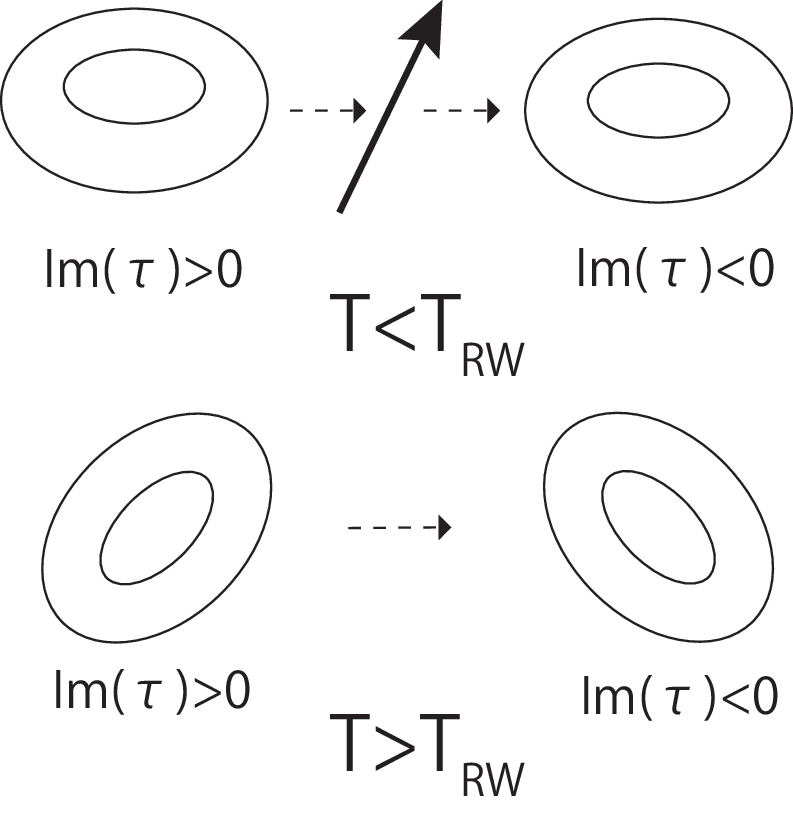}\\
\caption{The image of the RW transition based on the torus deformations. The bold arrow is a one-dimensional object.}
\label{F_RWtransition}
\end{figure}
%%%%%%%%%%%%%%%

%%%%%%%%%%%%%
\begin{figure}[b]
\centering
\includegraphics[width=0.4\textwidth]{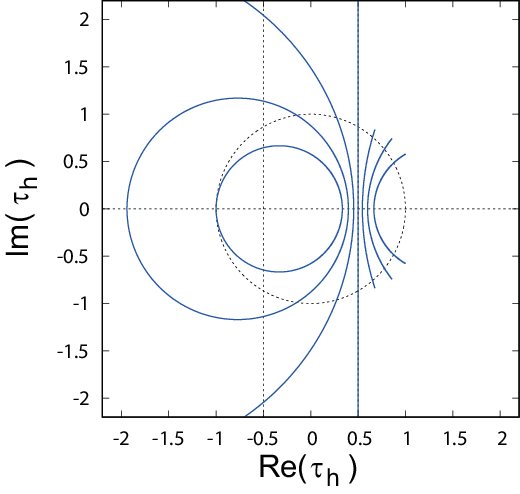}\\
\caption{The solid lines show $\tau_\mathrm{h} $ with $|\tau | =0.5,0.66,0.83,1,1.2,1.5$ and $2$.  
When $T\to T_{\rm RW}-0$ and $|\tau |=1$, $\tau_\mathrm{h}$ forms the line ${\mathrm Re}(\tau_\mathrm{h} )=0.5$. 
When $T\to T_{\rm RW}+0$ and $|\tau |=1$, $\tau_\mathrm{h}=0.5+iy~(y=-\sqrt{3}/2\sim \sqrt{3}/2 )$.  
When $T<T_{\mathrm RW}$ and $|\tau |<1$, $\tau_\mathrm{h}$ forms a circle. 
As $|\tau |$ increases, the circle approaches the line ${\mathrm Re}(\tau_\mathrm{h} )=0.5$. 
When $T>T_{\mathrm RW}$ and $|\tau |>1$, $\tau_\mathrm{h}$ forms an arc. 
As $|\tau |$ deceases, the arc approaches the line ${\mathrm Re}(\tau_\mathrm{h} )=0.5$. 
}
\label{F_tau_EVE}
\end{figure}
%%%%%%%%%%%%%%%

Figure~\ref{F_r095_l} shows the moduli $\tau$ and $\tau_{\rm f}$ when $T<T_{\rm RW}$ and $|\tau |=0.95$. 
 Again, the transformed $\tau_{\rm f}$ is symmetric with respect to the line ${\rm Re}(\tau)=0$.

When the order of the endpoint of the RW transition line is second-order, we can consider the limit $T \to T_{\rm RW}\pm 0$ and $|\tau |=1$. 
Figure~\ref{F_r10_l} shows the moduli $\tau$ and $\tau_{\rm f}$ when $T\to T_{\rm RW}-0$ and $|\tau |=1$.  
We see that $\tau$ is transformed into the boundary of the fundamental region. 
$\tau$ and $\tau_{\rm f}$ are symmetric with respect to the line ${\rm Re}(\tau )=0$.  
Note that the left boundary can be transformed into the right one by T transformation and vice versa.

Figure~\ref{F_r10_h} shows the moduli $\tau$ and $\tau_{\rm f}$ when $T\to T_{\rm RW} +0$ and $|\tau |=1$.  
$\tau$ is not symmetric with respect to the line ${\rm Re}(\tau )=0$, but 
$\tau_{\rm f}$ is, 
since the left boundary of the fundamental region can be transformed into the right one by the T transformation and vice versa.  
Also note that $\tau_{\rm f}$ in Fig.~\ref{F_r10_h} resembles that in Fig.~\ref{F_r10_l}, but the value of $\tau_{\rm f}(\theta \to \pi )$ is different. 
In this sense, the symmetry is broken even for $\tau_{\rm f}$ in Fig.~\ref{F_r10_h}. 
Also note that $\tau =(-0.5,\pm \sqrt{3}/2)$ is a fixed point of the modular transformation ${\cal{ST}}$.  
This transformation generates a discrete subgroup of the modular group, namely, ${\mathbb{Z}}_3$ group, since $({\cal{ST}})^3$ is an identity transformation.  
It is known that this property is related to the residual ${\mathbb{Z}}_3$-symmetry in modular flavor symmetry~\cite{Kikuchi:2022geu}.

Figure~\ref{F_r105_h} shows the moduli $\tau$ and $\tau_{\rm f}$ when $T > T_{\rm RW}$ and $|\tau |=1.05$.  
The transformed $\tau_{\rm f}$ resembles the one with $|\tau| =0.95$. 
This feature is originated in the fact that $\tau$ can be transformed into $-1/\tau$ by the S transformation. 
However, not only $\tau$ but $\tau_{\rm f}$ are not symmetric with respect to the line ${\rm Re}(\tau )=0$. 
The left-right symmetry is broken at high temperature. 

Figure~\ref{F_r20_h} shows the moduli $\tau$ and $\tau_{\rm f}$ when $T > T_{\rm RW}$ and $|\tau |=2$.  
Not only $\tau$, but also $\tau_{\rm f}$ are not symmetric with respect to the line ${\rm Re}(\tau )=0$. 
The transformed $\tau_{\rm f}$ resembles the one with $|\tau| =0.5$, but the left-right symmetry is broken.

In Fig.~\ref{F_RWtransition}, we summarize the image of the RW transition based on the torus deformations. 
When $T<T_{\rm RW}$, the torus shrinks to the one-dimensional object at $\theta =\pi$ smoothly, and then the object transforms into another torus $\mathrm{Im}(\tau )$ of which has an opposite sign, but discontinuity does not appear.
Since the modular transformation cannot change the sign of $\mathrm{Im}(\tau )$, the torus with a positive $\mathrm{Im}(\tau )$ cannot be smoothly transformed into the one with a negative $\mathrm{Im}(\tau )$ without shrinking of the torus. 
When, $T>T_{\rm RW}$, the torus changes discontinuously at $\theta =\pi$. 
The structure of the torus with modulus $\tau$ at high $T$ is similar to that with modulus $-1/\tau$ at low $T$ but the left-right symmetry is broken at high $T$.

Note that we have used the $\theta$-dependence of the phase of the Polyakov loop as a dynamical input. 
Although the structure of the torus is invariant under modular transformation, the dynamics itself is related to the details of the modulus, namely the structure of EoS. 
In Fig.~\ref{F_tau_EVE}, we show $\tau_\mathrm{h}$ given by Eq.~(\ref{n_b_hybrid_2}) in the hybrid model. 
When $T\to T_{\rm RW}-0$ and $|\tau |=1$, $\tau_\mathrm{h}$ forms the line $\mathrm{Re}(\tau_\mathrm{h} )=0.5$.  
In fact, it is easily shown 
%%%%%%%%%%%%%%
\begin{align}
\tau_\mathrm{h} 
 &={1\over{e^{-i\theta}+1}}
\nonumber\\
 &= 0.5+i{\sin{\theta}\over{2(\cos{\theta}+1)}},
\label{taud_line}
\end{align}
%%%%%%%%%%%%%%
Although it is not clear in the figure, $\tau_\mathrm{h} =0.5+iy$ with $y=-\sqrt{3}/2\sim \sqrt{3}/2$ when $T\to T_{\rm RW}+0$. 
Note that Eq.~(\ref{taud_line}) is also valid for this case if we replace $\theta$ by $\Delta \phi_{\mathrm b-q}$ in Eq.~(\ref{D_phase_high}). 
When $T<T_{\mathrm RW}$ and $|\tau |<1$, $\tau_\mathrm{h}$ forms a circle. 
It can be shown that 
%%%%%%%%%%%%%%%
\begin{eqnarray}
&&\tau_\mathrm{h}
={1\over{|\tau |^{-1}e^{-i\theta}+1}}
={1\over{2}}-s+x+iy~;
\nonumber\\
&&x={st+{1\over{2}}(|\tau |^2-1)\over{t}},~~~y={|\tau |\sin{\theta}\over{t}}, 
\nonumber\\
&&x^2+y^2=s^2-{1\over{4}}>0~; 
\nonumber\\
&& s=-{1\over{2}}{|\tau |^2+1\over{|\tau |^2-1}},~~~t=2|\tau |\cos{\theta}+1+|\tau |^2. 
\nonumber\\
\label{circle_taud}
\end{eqnarray}
%%%%%%%%%%%%%%%
Since $s^2-1/4$ is real positive constant which does not depend on $\theta$, $\tau_\mathrm{h}$ forms a circle.
The circle vanishes in the limit $|\tau |\to 0$.  
The center of the circle is given by 
%%%%%%%%%%%%%%%
\begin{eqnarray}
(x_0,y_0) = \left({1\over{2}}-s,0 \right).
\label{center_of_c}
\end{eqnarray}
%%%%%%%%%%%%%%%
As $|\tau |$ increases from 0, the right side of the circle approaches the line $\mathrm{Re}(\tau_\mathrm{h} )=0.5$, although $x_0$ moves the opposite direction. 
When $T>T_{\mathrm RW}$ and $|\tau |>1$, $\tau_\mathrm{h}$ forms an arc. 
Note that Eq.~(\ref{circle_taud}) is valid also for $|\tau |>1$ if we replace $\theta$ by $\Delta \phi_{\mathrm b-q}$ in Eq.~(\ref{D_phase_high}). 
As $|\tau |$ decreases from $\infty$, the arc approaches the line $\mathrm{Re}(\tau_\mathrm{h} )=0.5$, although $x_0$ moves in the opposite direction. 
This figure also indicates the phase transition at $|\tau |=1$.

It is also very interesting that the first equations of (\ref{taud_line}) and (\ref{circle_taud}) have the same $\theta$-dependence as the Fermi distribution function with complex chemical potential, and $\tau =|\tau |e^{i\theta}$ has the same one as the Boltzmann distribution, although the physical meanings of $\tau_\mathrm{h}$ and $\tau$ are different from those of the distribution functions. Hence, these equations also indicate that the Boltzmann distribution can be transformed into the Fermi distribution by the modular transformation with the matrix product $\cal{TST}$. 
When $T \to 0$, the phase transition occurs at $|\tau |=e^{(\mu_\mathrm{FR}-m_\mathrm{F})/T}=1$ and $p_\mathrm{F}=0$ where $m_\mathrm{F}$, $\mu_\mathrm{FR}$ and $p_\mathrm{F}$ are the fermion mass, the real part of the fermion chemical potential, and the Fermi momentum, respectively. 
That is nothing but the formation of Fermi surface.

%\clearpage

%%%%%%%%%%%%%%%%%%%%%%%%%%%%%%%%%
%%%%%%%%%%%%%%%%%%%%%%%%%%%%%%%%%
%%%%%%%%%%%%%%%%%%%%%%%%%%%%%%%%%
\section{Summary}
\label{summary}
%%%%%%%%%%%%%%%%%%%%%%%%%%%%%%%%%
%%%%%%%%%%%%%%%%%%%%%%%%%%%%%%%%%
%%%%%%%%%%%%%%%%%%%%%%%%%%%%%%%%%

In summary, we have reformulated the recently proposed hadron-quark hybrid model in the framework of modular transformation when the imaginary chemical potential $\mu =i\theta T$ is introduced. 
We can consider the torus, which is characterized by the number densities of baryons (antibaryons) and quarks (antiquarks). 
When $T<T_{\rm RW}$, the torus shrinks to the one-dimensional objects smoothly at $\theta =\pi$, and then transformed into the another torus. 
When $T>T_{\rm RW}$, the torus changes discontinuously at $\theta =\pi$. 
We also calculated the modulus of the torus and transformed it into the fundamental region. 
The transformed modulus $\tau_{\rm f}$ at high temperature resembles that at low temperature, but the left-right symmetry is broken at high temperature. 
The modulus $\tau_\mathrm{h}$ of the hybrid model itself also indicates the phase transition at the RW transition point.  

In this paper, we have restricted our discussions on the region where temperature $T$ is above or just below $T_{\rm RW}$. 
When $T$ is much smaller than $T_{\rm RW}$, the baryon volume approaches the constant value $v_0$ and $n_{\rm b}$ is small. 
Then, the hybrid model reduces to the ordinary hadron resonance gas model~\cite{Oshima:2023bip}. 

We also remark that, in this paper, the $\theta$-dependence of the phase $\phi$ of the Polyakov loop $\Phi$ was used as a dynamical input.  
Usually, the absolute value of $\Phi$ is used to classify the phases of QCD. 
However, the phase $\phi$ is also a very important quantity. 
For example, in Refs.~\cite{Hirakida:2018bkf,Kashiwa:2021ctc},  the $\phi$-distribution in the phase space is used to analyze the confinement-deconfinement transition in the framework of persistent homology analyses. 

In this paper, we only performed qualitative analyses of the hybrid model using the $\theta$-dependence of the phase of the Polyakov loop obtained by the lattice QCD and some phenomenological assumptions. 
To perform the quantitative analyses, we need to calculate the thermodynamic potential of the hybrid model and determine the solutions of the order parameters such as the Polyakov loop and the chiral condensates within the model.  
Such studies are in progress.   

\begin{acknowledgments}
The authors thank H. Aoki for useful discussions on the modular transformations.
This work is supported in part by Grants-in-Aid for Scientific Research from JSPS (No. JP22H05112).
\end{acknowledgments}

\appendix

\section{Lattice and torus}
\label{Ltorus}

In this appendix, we brief review the relation between the lattice, the torus and the modular transformation.  
For a more detailed description, see, eg, the chapter 8 in Ref.~ \cite{Nakahara:2003nw}. 
 
\begin{figure}[t]
\centering
\includegraphics[width=0.28\textwidth]{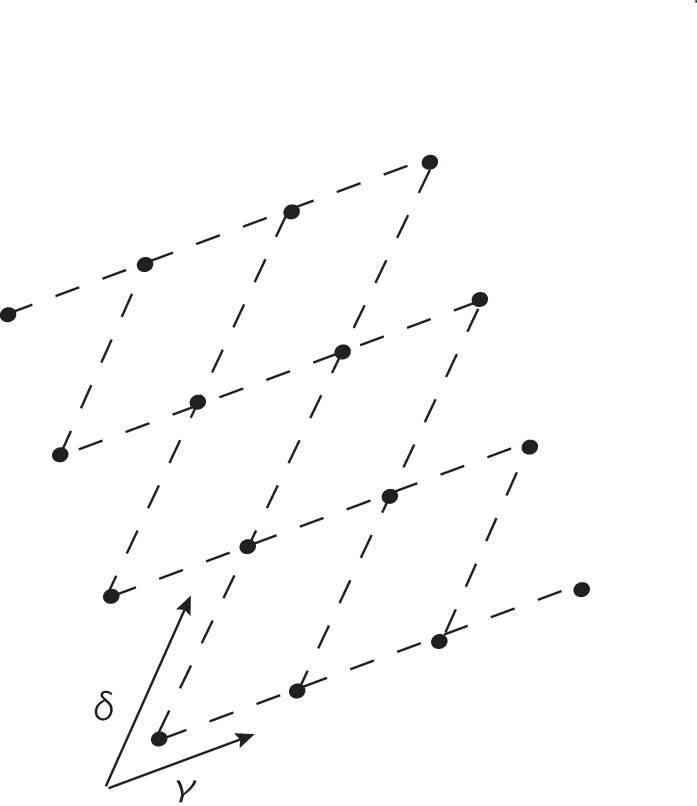}\\

\caption{The lattice $\Lambda$ in the complex plane. 
The dots are lattice points. 
}
\label{Lattice_1}
\end{figure}

When $\gamma,\delta \in {\mathbb{C}}-\{ 0 \}$ and $\tau ={\delta\over{\gamma}} \notin {\mathbb{R}}$, 

%%%%%%%%%%%%%%%%%%%
\begin{eqnarray}
\Lambda =\{ m\gamma + n\delta \}~(n,m \in {\mathbb{Z}})
\label{A_lattice_1}
\end{eqnarray}
%%%%%%%%%%%%%%%%%%%
is called a "lattice" in the complex plane. (See Fig.~\ref{Lattice_1} )

%%%%%%%%%%%%%%
%%%%%%%%%%%%%%

\begin{figure}[t]
\centering
\includegraphics[width=0.20\textwidth]{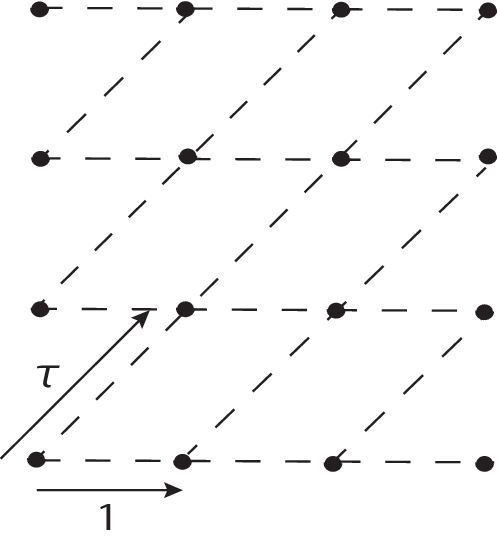}\\

\caption{The lattice $\Lambda_\tau$ in the complex plane. 
The dots are lattice points.  
}
\label{Lattice_tau}
\end{figure}

Equivalently, 
%%%%%%%%%%%%%%%%%%%%
\begin{eqnarray}
\Lambda_\tau =\{m+n\tau ~(n,m \in {\mathbb{Z}})\}, 
\label{A_lattice_2}
\end{eqnarray}
%%%%%%%%%%%%%%
is also a lattice.  See Fig. ~\ref{Lattice_tau}.  

%%%%%%%%%%%%%%
%%%%%%%%%%%%%%

The lattice $\Lambda^\prime$ is called 
"similar" to the lattice $\Lambda$ when
$\alpha\Lambda^\prime =\Lambda$ is satisfied
for a nonzero $\alpha \in {\mathbb{C}}$. 
It is easily shown that $\Lambda_\tau$ is similar to $\Lambda$. 
For the modular transformation (\ref{modular}), it can be also shown that  
%%%%%%%%%%%%%%
\begin{eqnarray}
\Lambda_{\tau^\prime }=(c\tau +d)^{-1}\Lambda_\tau. 
\label{A_similar}
\end{eqnarray}
%%%%%%%%%%%%%%%
Hence, $\Lambda_{\tau^\prime} $ is similar to $\Lambda_\tau$. 

By identifying the opposite edges of the unit lattice, we can construct the torus $T^2$ the complex structure of which is characterized by $\tau$. 
Then, $\tau$ is called a "modulus" of the torus. 
The torus is the quotient space $\mathbb{C}/\Lambda$. 
When the lattice $\Lambda^\prime$ is similar to the lattice $\Lambda$, the complex structure of the torus ${T^2}^\prime $ is the same as that of $T^2$. Hence, the complex structure of the torus is invariant under the modular transformation. 
By considering the torus, we can extract the topological property which is invariant under the modular transformation.

%\clearpage

\bibliography{ref.bib}

\end{document}